\documentstyle[prl,aps,multicol]{revtex}
\input{epsf}
\begin{document}
\draft
\title{Emergence of Quantum Ergodicity in Rough Billiards}

\author{Klaus M. Frahm and Dima L. Shepelyansky$^{*}$}

\address {Laboratoire de Physique Quantique, UMR C5626 du CNRS, 
Universit\'e Paul Sabatier, F-31062 Toulouse Cedex 4, France}

\date{14 February 1997}

\maketitle

\begin{abstract}
By analytical mapping of the eigenvalue problem in rough billiards 
on to a band random matrix model a new regime of Wigner ergodicity 
is found. There the eigenstates are extended over 
the whole energy surface but have a strongly peaked structure. The 
results of numerical simulations and implications 
for level statistics are also discussed. 
\end{abstract}


\begin{multicols}{2}
\narrowtext

\begin{figure}
\epsfxsize=3in
\epsfysize=5in
\epsffile{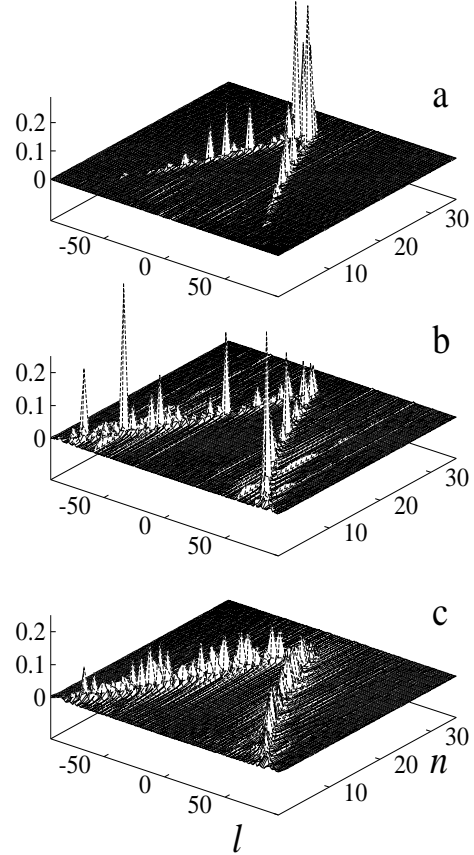}
\vglue 0.2cm
\caption{Transition from localization to Shnirelman ergodicity on 
energy surface for level number $N\approx 2250$, 
$l_{max}\approx 95$ and $M=20$; 
shown are the absolute amplitudes $|C^{(\alpha)}_{nl}|$ 
of one eigenstate: (a) localization for $D(l_r=0)=20$; 
(b) Wigner ergodicity for $D=80$; (c) Shnirelman ergodicity for $D=1000$
(see text).
}
\label{fig1}
\end{figure}

In 1974, Shnirelman \cite{shnirel1} proved a theorem according to which 
quantum eigenstates in chaotic billiards become ergodic for sufficiently 
high level numbers. Later it was demonstrated \cite{bohigas,leshouches} 
that in this regime the level spacing statistics $p(s)$ 
is well described by random matrix 
theory \cite{mehta}. However, one can ask the question how this 
quantum ergodicity emerges with increasing level number $N$? This question 
becomes especially important in the light of recent results 
\cite{fausto,rough} for diffusive billiards where the time of 
classical ergodicity $\tau_D$ due to diffusion on the energy surface 
is much larger than the collision time with the boundary $\tau_b$. 
In such a situation 
quantum localization on the energy surface may break classical 
ergodicity eliminating the level repulsion in $p(s)$. The 
investigation of rough billiards \cite{rough} showed that this change 
of $p(s)$ happens when the localization length $\ell$ in the angular momentum 
$l$-space becomes smaller than the size of the energy surface characterized 
by the maximal $l=l_{max}$ at given energy ($\ell < l_{max}$). For 
$\ell>l_{max}$ the eigenfunctions are extended over the whole surface but 
as we will see they are not necessarily ergodic (Fig. \ref{fig1}). 

In this situation which we will furthermore call {\em Wigner ergodicity} 
the eigenstates are composed of rare strong peaks distributed on the 
{\em whole} energy surface. Such a case is very different from the 
{\em Shnirelman ergodicity} where the eigenstates are uniformly 
distributed. The usual scenario of ergodicity breaking was 
based on the image of quantum localization {\em along} the 
energy surface \cite{chirikov}. Here we show that the transition between 
localized and Shnirelman ergodic states can pass trough an intermediate 
phase of Wigner ergodicity. Our description and understanding of this 
phase is based on the mapping of the billiard problem with weakly 
rough (random) boundary on to a superimposed 
band random matrix (SBRM). This model is characterized by strongly fluctuating 
diagonal elements corresponding to a preferential basis of the 
unperturbed problem. Recently such type of matrices was studied in 
the context of the problem of particle interaction in disordered 
systems \cite{dima1,dima2,fyodorov,klaus1}. There it was found that 
the eigenstates can be extended over the whole matrix size while having 
a very peaked structure. The origin of this behavior is due to the 
Breit-Wigner form \cite{wigner} of the local density of states according 
to which 
only unperturbed states in a small energy interval 
$\Gamma_E$ contribute to the final eigenstate. 
\vspace{-1cm}
\noindent

Recent optical experiments with micrometer size droplets initiated new 
theoretical investigations of weakly deformed circular billiards 
\cite{stone1}. In this case the ray dynamics becomes chaotic 
leading to a strong directionality of light emission \cite{stone2}. Here 
we will consider another type of a weakly deformed circle \cite{rough}, namely 
we chose a random elastic boundary deformation which can be represented by
$R(\theta)=R_0+\Delta R(\theta)$ with $\Delta R(\theta)/R_0=
\mbox{Re}\ \sum_{m=2}^M \gamma_m\,e^{i m \theta}$ where $\gamma_m$ 
are random complex coefficients and $M$ is large but finite. 
This type of deformation seems to be very generic and may appear in 
numerous different physical situations \cite{rough}. In small droplets 
such boundary perturbations may be created by temperature induced surface 
waves. We will restrict ourself to the case of weak surface roughness 
given by $\kappa(\theta)=(d R/d\theta)/R_0\ll 1$ and all $\gamma_m$ 
being of the same order of magnitude. Then we have for the angle average 
$\tilde \kappa^2=\langle \kappa^2(\theta)\rangle_\theta \sim 
M^2 (\Delta R/R_0)^2$. 

\begin{figure}
\epsfxsize=3in
\epsffile{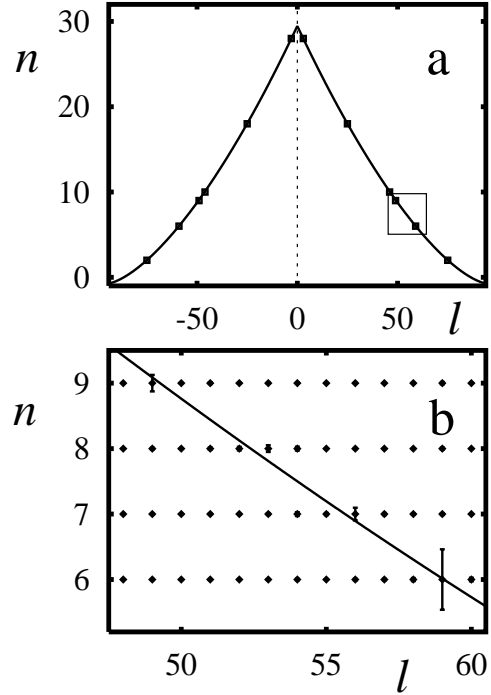}
\vglue 0.2cm
\caption{(a) Main peaks of eigenstate in Fig. 1b 
(squares for $|C_{nl}^{(\alpha)}| \ge 0.1$) shown on the energy surface 
${\cal H}(n,l)=E_\alpha$ (see text); 
(b) rescaled part of (a): diamonds show the integer $(n,l)$-lattice, 
the errorbar size is $2 |C_{nl}^{(\alpha)}|$. }
\label{fig2}
\end{figure}

In such a billiard the dynamics is diffusive in orbital momentum 
due to collisions with the rough boundary provided 
$\tilde\kappa$ is above the chaos border $\kappa_c \sim M^{-5/2}$ 
\cite{rough}. The diffusion constant is determined by the average 
change of orbital momentum per collision being $D=<(\Delta l)^2>
=4\,(l_{max}^2-l_r^2)\, \tilde \kappa^2$. This $D$ is the local 
diffusion rate for $l$ close to $l_r$. The 
quantum interference leads to localization of this diffusion 
with the length $\ell=D$ for $M < \ell < l_{max}$ while for 
$\ell > l_{max}$ the eigenstates are extended over the energy surface 
\cite{rough}. The transition between these two regimes is illustrated 
in Fig. \ref{fig1}. Here we present the absolute values 
of eigenfunction amplitudes $C^{(\alpha)}_{nl}$ in the eigenbasis $|nl>$ of 
circular billiard as a function of unperturbed radial and orbital quantum
numbers $n$, $l$ with $\alpha$ marking the eigenenergy $E_\alpha$. 
For small roughness $\tilde\kappa$ (or $D$) the states are 
exponentially localized 
(Fig. \ref{fig1}a) while for large $\tilde\kappa$ they are homogeneously 
distributed (Fig. \ref{fig1}c) on the energy surface. The case of Fig. 
\ref{fig1}b corresponds to an unusual regime of Wigner ergodicity where the 
eigenstate is extended over the surface but is composed of rare strong 
peaks. The positions of these peaks on the energy surface of the circular 
billiard $E={\cal H}(n,l)$ are shown in Fig. \ref{fig2}a. The equation 
of the surface, projected on the action plane $(n,l)$, can be found from 
the Bohr-Sommerfeld quantization $\mu_l(E)=\sqrt{l_{max}^2-l^2} - 
l \arctan(l^{-1} \sqrt{l_{max}^2-l^2})+\pi/4=\pi(n+1)$ where $l_{max}^2=4 N=
2m R_0^2 E/\hbar^2=k^2 R_0^2$ with $k$ being the wave number. 
A part of the surface is shown in more detail in Fig. \ref{fig2}b. 
Here it is clearly seen that the peaks are large for those integer 
$n$, $l$ which are close to the line ${\cal H}(n,l)=E_\alpha$. 
Our understanding of the fact that not all integer values of the 
$(n,l)$-lattice near to this line are populated is based on the 
concept of Breit-Wigner structure of eigenstates described below. 

According to Refs. \cite{rough,klaus2} the internal scattering at 
the rough boundary can be described by the $S$-matrix
\begin{equation}
\label{quantum_map}
S_{l\tilde l}(E)  =  e^{i\mu_l(E)} <l|e^{i\,V(\theta)}|\tilde l>
e^{i\mu_{\tilde l}(E)}\ ,\\
\end{equation}
with $\mu_l(E)$ being the scattering phases of the circle and 
$V(\theta)= 2\sqrt{l_{max}^2-l_r^2}
\ \Delta R(\theta)/R_0$.
This quantum rough map \cite{rough} is defined with respect to 
amplitudes $a_l$ in the wave function expansion 
$\psi(r,\theta)=B \sum_l a_l\,J_{|l|}(kr)\,e^{il\theta}$ 
with Bessel functions $J_l$ and $B$ being a normalization constant. 
The $S$-matrix gives a local unitary description for $l$ close to a 
resonant value $l_r$. The eigenvalue equation reads $\sum_{\tilde l} 
S_{l,\tilde l}(E_\alpha)\,a_{\tilde l}^{(\alpha)}=a_l^{(\alpha)}$ 
so that the eigenvalues $E_\alpha$ are determined by 
$\det[1-S(E_\alpha)]=0$. For $V=0$, we recover the Bohr-Sommerfeld 
quantization for eigenvalues $E_{nl}$ of the ideal circle.

The semiclassical regime of ray dynamics corresponds to the limit 
$V\gg 1$ where the $\theta$-integral can be evaluated in a
saddle point approximation giving the classical limit of quantum rough 
map \cite{rough}. 
Here we are interested in a different regime where $V<1$ corresponding 
to $D<M^2$. There, by the mapping on an effective solid state Hamiltonian 
$H_{eff}$ introduced by Fishman, Grempel, and Prange \cite{fishman}, the 
equation for eigenstates takes the form 
\begin{equation}
\label{tan_eq}
\tan[\mu_l(E_\alpha)]\, a_l^{(\alpha)} + \frac{1}{2}\,\sum_{\tilde l} 
<l|V|\tilde l>\, a_{\tilde l}^{(\alpha)}=0\ .
\end{equation}
In this way the eigenvalue equation is reduced to a solid state problem 
with $2M$ coupled sites. The $H_{eff}$-matrix is of SBRM type with 
strongly fluctuating diagonal elements produced by scattering phases 
$\mu_l$. The investigations of such matrices \cite{dima2,fyodorov,klaus1} 
showed that the local density of states has the Breit-Wigner width 
given by the Fermi golden rule 
$\Gamma_\mu=2\pi \rho_\mu <(V(\theta)/2)^2>\approx 3D/(2M^2)$ 
where $\rho_\mu=1/\pi$ is the density of diagonal elements and we used 
the relation between phase average of $V^2(\theta)$ and $D$. This 
expression is valid \cite{note1} 
when $\Gamma_\mu$ exceeds the mean level spacing 
($\sim 1/M$) in the band width $M$. In the opposite limit $\Gamma_\mu M < 1$
the eigenstates are given by standard perturbation theory. Together 
with the condition $V<1$ we find that the Breit-Wigner regime exists 
for $M < D < M^2$ near zero energy of $H_{eff}$. 
In this regime the localization length is $\ell=D$ 
\cite{rough,dima1,dima2,fyodorov,klaus1,klaus2}. 
However, the Breit-Wigner 
structure remains in both localized ($\ell<l_{max}$) and delocalized 
($\ell>l_{max}$) cases if $M < D < M^2$. Therefore for 
$l_{max} < D < M^2$ the states are extended but only $l$ 
with $|\tan(\mu_l)|<\Gamma_\mu<1$ are mixed leading to a 
peaked structure of eigenstates \cite{note2}. 
The fraction of peaks in $\max(\ell,l_{max})$ is $\Gamma_\mu$. 

\begin{figure}
\epsfxsize=2.7in
\epsffile{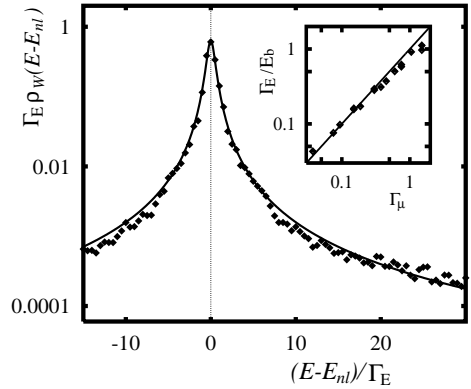}
\vglue 0.2cm
\caption{Breit-Wigner distribution for eigenstates of rough billiard 
(diamonds) 
with the parameters of Fig. 1b (5 eigenstates for each of 10 roughness 
realizations are used). Full curve gives the distribution (\protect\ref{bw}) 
with the theoretical $\Gamma_E$ value (\protect\ref{ge}), $\Gamma_E/
E_b=\Gamma_\mu=0.3$. 
Insert shows the variation of $\Gamma_E/E_b$ (diamonds) as a function of 
$\Gamma_\mu$ for parameter range $10\le M\le 40$ and $M\le D\le M^2$ 
($E_b$ and $\Gamma_\mu$ are taken at $l=0$). The theory (\protect\ref{ge}) 
is shown by straight line.
}
\label{fig3}
\end{figure}

The above properties of scattering amplitudes $a_l^{(\alpha)}$ 
allow to understand the behavior of eigenfunction coefficients 
$C_{nl}^{(\alpha)}=<\psi_\alpha|nl>$. For this one has to compute the 
expansion $J_l(k_\alpha\,r)e^{il\theta}$ in terms of $|nl>$. Since 
$\Delta R\ll R_0$ the angular and radial integrals factorize and 
can be evaluated using the radial eigenvalue equation and the 
semiclassical expression for $J_l(kr)$. As a result we obtain
\begin{equation}
\label{c_rel}
C_{nl}^{(\alpha)}  \approx \tilde B  a_l^{(\alpha)}\ (l_{max}^2-l^2)^{1/4}\ 
\frac{\sin \Delta\mu}{\Delta\mu}\ ,
\end{equation}
with $\Delta\mu = \mu_l(E_\alpha)-\mu_l(E_{nl})\approx 
(E_\alpha-E_{nl})/E_b$ and $E_b=dE/d\mu_l(E)=\hbar^2 l_{max}^2/(m R_0^2
\sqrt{l_{max}^2-l^2}\,)=2\hbar/\tau_b$ being the energy scale 
related to the ballistic collision time $\tau_b$, $\tilde B$ is 
a normalization constant.  
The amplitudes $C_{nl}^{(\alpha)}$ determine the local 
density of states by
\begin{equation}
\label{loc_dens_def}
\rho_{W}(E-E_{nl})=\Big\langle \sum_\alpha \delta(E-E_\alpha)\,
|C_{nl}^{(\alpha)}|^2\Big\rangle
\end{equation}
The averaging is performed with respect to different roughness 
realizations and/or over a sufficiently large energy interval. 
Due to the Breit-Wigner distribution for $\tan(\mu_l)$ in (\ref{tan_eq}) 
we obtain 
\begin{equation}
\label{bw}
\rho_{BW}(E-E_{nl})=\frac{1}{\pi}\,\frac{\Gamma_E/2}{
(E-E_{nl})^2+\Gamma_E^2/4}
\end{equation}
with 
\begin{equation}
\label{ge}
\Gamma_E=E_b\,\Gamma_\mu=E_b \frac{N}{N_W}\,\left(1-
\frac{l^2}{l_{max}^2}
\right)\ ,\ N_W=\frac{M^2}{24 \tilde\kappa^2}\ .
\end{equation}
The equations (\ref{bw}), (\ref{ge}) are valid for $\Delta E<E_B$ 
($\Delta\mu < 1$) and $\Gamma_\mu<1$ or $N<N_W$. Here $N_W$ is the 
border of Breit-Wigner regime in level number $N$. We remind that 
the eigenstates are localized for $N<N_e=1/(64\tilde\kappa^4)$ 
corresponding to $\ell<l_{max}$ \cite{rough}. As a result the 
Breit-Wigner structure can exist both in the localized and delocalized 
cases. An example of Breit-Wigner distribution is shown in Fig. \ref{fig3}. 
Our numerical data confirm the theoretical expression (\ref{ge})  
for variation of $\Gamma_\mu$ by more than one order of magnitude (inset).

For $N>N_W$ the kick amplitude $V$ in (\ref{quantum_map}) is larger than 1 
and the mapping on to equation (\ref{tan_eq}) is not valid. In this case 
the scattering phases (eigenphases of $S$) are homogeneously distributed 
in the interval $(0,2\pi)$. If in addition $N>N_e$ then as in the case 
of kicked rotator (see B. V. Chirikov in \cite{leshouches}) the amplitudes 
$a_l$ are homogeneous in $l$-space with $|a_l|^2\approx 1/(2\l_{max})$. 
Using Eq. (\ref{c_rel}), we obtain that the local density of states is 
given by
\begin{equation}
\label{loc_dens_erg}
\rho_{W}(E-E_{nl})=\frac{E_b}{\pi}\,
\frac{\sin^2 [(E-E_{nl})/E_b]}{(E-E_{nl})^2}\ .
\end{equation}
This density is normalized to one and as a result the probability 
$|C_{nl}^{(\alpha)}|^2$ is ergodically distributed along the 
energy surface shown in Fig. \ref{fig2}a. This is the regime of Shnirelman 
ergodicity which emerges for $N>\max(N_W,\,N_e)$. For fixed roughness 
$\tilde\kappa>\kappa_{EW}=\sqrt{6}/4M$ we have $N_W>N_e$ and the transition 
to Shnirelman ergodicity with the increasing level number $N$ crosses the 
region of Wigner ergodicity for which an eigenfunction is ergodic only inside 
the Breit-Wigner width $\Gamma_E<E_b$. In the opposite case $\kappa_c<
\tilde\kappa<\kappa_{EW}$ the Shnirelman ergodicity emerges directly from 
the localized phase (the Breit-Wigner regime exists only in the localized 
phase). The averaging of equation (\ref{loc_dens_erg})
over different $l$-values gives the distribution (\ref{bw}) with 
$\Gamma_E\sim E_b$. 
This explains why also for the case $\Gamma_\mu<1$ in Fig. 
\ref{fig3} the distribution (\ref{bw}) remains valid even for 
$\Delta E>E_b$ (note that $\pi E_b/\Gamma_E\approx 10$). 

\begin{figure}
\epsfxsize=3in
\epsffile{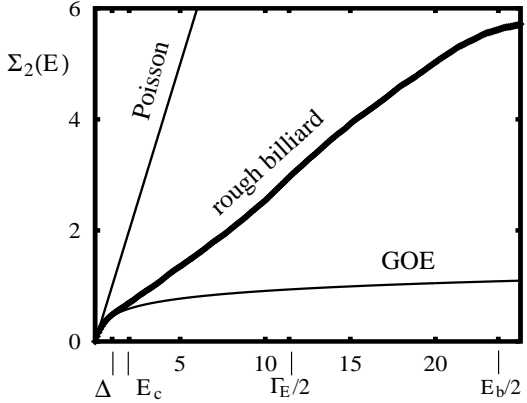}
\vglue 0.2cm
\caption{Dependence of number variance $\Sigma_2(E)$ on energy for 
rough billiard compared to Poisson and GOE 
for $M=50$, $D=800$, $l_{max}\approx 95$; 6 roughness realizations 
in the interval $2150 <N< 2350$ are used for the average. The 
energy scales are also shown in units of level spacing $\Delta$. 
}
\label{fig4}
\end{figure}

The above analysis shows that in the regime of Wigner ergodicity 
there are four relevant energy scales:
level spacing $\Delta$, Thouless energy for diffusion in $l$-space 
$E_c=\hbar D/(l_{max}^2 \tau_b)$, the Breit-Wigner width $\Gamma_E
=3\hbar D/(M^2\tau_b)$ and bouncing energy $E_b=2\hbar/\tau_b$ 
which are ordered as $\Delta < E_c < \Gamma_E < E_b$. These scales 
should appear in the level statistics namely for the number variance 
$\Sigma_2(E)$ \cite{leshouches,mehta}. For $E<E_c$ as usual we expect 
gaussian orthogonal ensemble (GOE) statistics to be valid while in the 
interval $E_c<E <\Gamma_E/2$ the behavior should be modified, due to 
the diffusive dynamics \cite{altshuler}, being $\Sigma_2(E)\sim 
(E/E_c)^{1/2}$. The first investigations of the regime with 
$\Gamma_E/2 < E < E_b/2$ for SBRM were done only recently \cite{dietmar}. 
They showed that level rigidity is strongly suppressed with a nearly 
linear energy behavior in $\Sigma_2(E)$ due to disappearance of 
correlations between levels with energy differences larger than 
$\Gamma_E$. However, such local characteristics as $p(s)$ 
are still described by GOE if $E_c\gg \Delta$. Our numerical data 
qualitatively confirm these expectations (see Fig. \ref{fig4}), however, 
quantitative numerical and analytical verifications are still required. 
In Fig. \ref{fig4} the above energy scales are not 
separated by strong inequalities but parametrically it is possible 
to have them. In this unusual 
regime it would be interesting to study other physical properties. 
We mention for example the frequency dependence of dielectrical response 
\cite{gorkov} which should be sensitive to the above energy scales. 

In conclusion, we studied the parameter dependence of the quantum 
energy surface width in rough billiards. In the limiting case of Shnirelman 
ergodicity with high level numbers, this width is determined by the 
typical frequency of collisions with the boundary ($\Gamma_E\sim E_b$). 
This means that {\it all} integer points on the $(n,l)$-lattice of 
quantum numbers with a distance $\Delta l=\Delta n \approx 1$ from 
the energy line $E_\alpha={\cal H}(n,l)$ are 
occupied by {\it one} eigenfunction $\psi_\alpha$ ($N>N_W$ and $N>N_e$). 
We have found a new regime of Wigner ergodicity where $\Gamma_E\ll E_b$ 
so that only points with $\Delta l=\Delta n \le \Gamma_\mu\ll 1$
contribute to $\psi_\alpha$. As a result a lot of holes appear in the 
energy surface and $\psi_\alpha$ has a strongly peaked structure 
on $(n,l)$-lattice. However, at the same time all states in the Breit-Wigner 
width $\Gamma_E$ are populated so that the eigenfunction is ergodic 
inside this energy band. It would be interesting to understand if 
the arithmetical properties of the lattice will play an important role.

\end{multicols}


\begin{thebibliography}{99}
\bibitem{note1} The situation here is a bit different from 
	\cite{dima2,fyodorov,klaus1} due to 
	translational invariance of $<l|V|\tilde l>$ in $l$. But this 
	does not change the result which can be derived exactly 
	for random phases $\mu_l$ using the Lloyd model fluctuations.
\bibitem{note2} Note that for the case studied in \cite{rough} with 
	$\gamma_m\sim 1/m$~this Breit-Wigner regime 
	does not exist, since there $D\sim M \langle V^2\rangle$ instead of 
	$D\sim M^2 \langle V^2\rangle$ for $\gamma_m\sim const$.
\end{thebibliography}
\end{document}